\documentclass[letter]{aa}

\usepackage{natbib}
\usepackage{graphicx}
\usepackage{txfonts}
\usepackage{color}

\newcommand{\msol}{$M_{\odot}$}

\newcommand{\lboo}{$\lambda\,$Bo\"{o}}
\newcommand{\mjup}{M$_{\rm Jup}$}

\newcommand{\revone}[1]{#1}
\newcommand{\revtwo}[1]{#1}

\begin{document}
\title{Fingerprints of giant planets in the photospheres of Herbig stars}

\author{
	M. Kama\inst{1}
	\and
	C.P. Folsom\inst{2,3}
	\and
	P. Pinilla\inst{1}
}

\institute{
	Leiden Observatory, P.O. Box 9513, NL-2300 RA, Leiden, The Netherlands, \email{mkama@strw.leidenuniv.nl}
	\and
	Universit\'{e} de Grenoble Alpes, IPAG, F-38000 Grenoble, France
	\and
	CNRS, IPAG, F-38000 Grenoble, France
}

\date{}

\abstract{
Around 2\% of all A stars have photospheres depleted in refractory elements. This is hypothesized to arise from a preferential accretion of gas rather than dust, but the specific processes and the origin of the material -- circum- or interstellar -- are not known. The same depletion is seen in 30\% of young, disk-hosting Herbig~Ae/Be stars. We investigate whether the chemical peculiarity originates in a circumstellar disk. Using a sample of systems for which both the stellar abundances and the protoplanetary disk structure are known, we find that stars hosting warm, flaring group I disks typically have Fe, Mg and Si depletions of 0.5 dex compared to the solar-like abundances of stars hosting cold, flat group II disks. The volatile, C and O, abundances in both sets are identical. Group I disks are generally transitional, having radial cavities depleted in millimetre-sized dust grains, while those of group II are usually not. Thus we propose that the depletion of heavy elements emerges as Jupiter-like planets block the accretion of part of the dust, while gas continues to flow towards the central star. We calculate gas to dust ratios for the accreted material and find values consistent with models of disk clearing by planets. Our results suggest that giant planets of ${\sim}0.1$ to $10\,$M$_{\rm Jup}$ are hiding in at least 30\% of Herbig~Ae/Be disks.
}
   \keywords{accretion, accretion disks; protoplanetary disks; planet-disk interactions; stars: variables: T Tauri, Herbig Ae/Be; stars: abundances; stars: chemically peculiar}
   \maketitle

\section{Introduction}

Chemically peculiar \lboo\ stars have a solar-like photospheric abundance of volatile elements (C, N, O) but are underabundant by up to a factor of $100$ in refractory elements (e.g., Fe, Mg, Si). The peculiarity affects up to $2\,$\% of all late B through early F stars \citep[e.g.,][]{GrayCorbally1998, Paunzen2001}. Among the young Herbig~Ae/Be stars, which host protoplanetary disks, the fraction is at least $33\,$\% \citep{Folsometal2012}. 

Stars of spectral type late B through early F have radiative envelopes, and consequently their photosphere is mixed with the interior on timescales ${\sim}1\,$Myr. This enables accreting material to dominate the photospheric composition for accretion rates as low as $10^{-13}\,$\msol$\,$yr$^{-1}$ \citep{TurcotteCharbonneau1993, Turcotte2002}. As the \lboo\ depletion pattern matches that of diffuse interstellar gas, it has been linked to the accretion of material from the circum- or interstellar medium, with radiation pressure providing the mechanism that keeps dust from accreting while gas streams onto the star \citep{VennLambert1990, KampPaunzen2002, MartinezGalarzaetal2009}. Since radiation pressure should affect all systems, it has been unclear why only a fraction of intermediate-mass stars display the \lboo\ peculiarity.

Herbig~Ae/Be stars are young B, A and F stars which host a protoplanetary disk and are accreting \citep{Herbig1960, WatersWaelkens1998}. Based on their spectral energy distributions, the disks are categorized as ``group~II'' (cold, flat) and ``group~I'' \citep[warm, flaring;][]{Meeusetal2001}. A further distinction is made between disks showing emission from warm, small silicate grains (II, Ia) and those not showing such features (Ib). Initially, group~I was thought to evolve to group~II as dust particles grew and settled \citep{Meeusetal2001, DullemondDominik2004a, DullemondDominik2004b}. The emerging view from recent debates is that group~I disks are transitional, i.e., have radial cavities or gaps strongly depleted in millimetre-sized dust and perhaps gas \citep[e.g.,][]{Keaneetal2014, Maaskantetal2014, BanzattiPontoppidan2015, Menuetal2015, vanderPlasetal2015}. As yet, no group II source is known to be a transition disk (TD), while all resolved TDs are group I sources.

Cavities and gaps in transitional disks may arise due to dust being trapped in the outer disk by, e.g., giant planets \citep{Whipple1972, BargeSommeria1995, Riceetal2006, Pinillaetal2012b, Zhuetal2012}, by gas and dust clearing due to photoevaporation \citep{Alexanderetal2006a, Alexanderetal2006b, Alexanderetal2014, Owenetal2011, Owenetal2012a}, or by the accumulation of gas and dust at the outer edge of a region with low ionization \citep{Regalyetal2012, Flocketal2015}. The dominant mechanism should determine the volatile (gas) to refractory (dust) elemental ratios in the accretion columns that channel material onto the stellar photosphere. Most Herbig~Ae/Be stars accrete at rates comparable to, or higher than, T~Tauri stars \citep[$10^{-9}$--$10^{-6}\,$\msol$\,$yr$^{-1}$;][]{GarciaLopezetal2006, DonehewBrittain2011, Pogodinetal2012}. We investigate whether stellar abundances correlate with disk properties, as might be expected from the above.

\section{Data}

\revone{Our sample, presented in Table~\ref{tab:systems}, consists of the overlap between the stellar abundance surveys and the compilation of disks described below. Metallicities or disk classifications for a few individual sources were added from the literature, but in this regard the sample is certainly not complete.}

\subsection{The elemental abundances}
\begin{figure}[!ht]
\includegraphics[clip=,width=1.0\columnwidth]{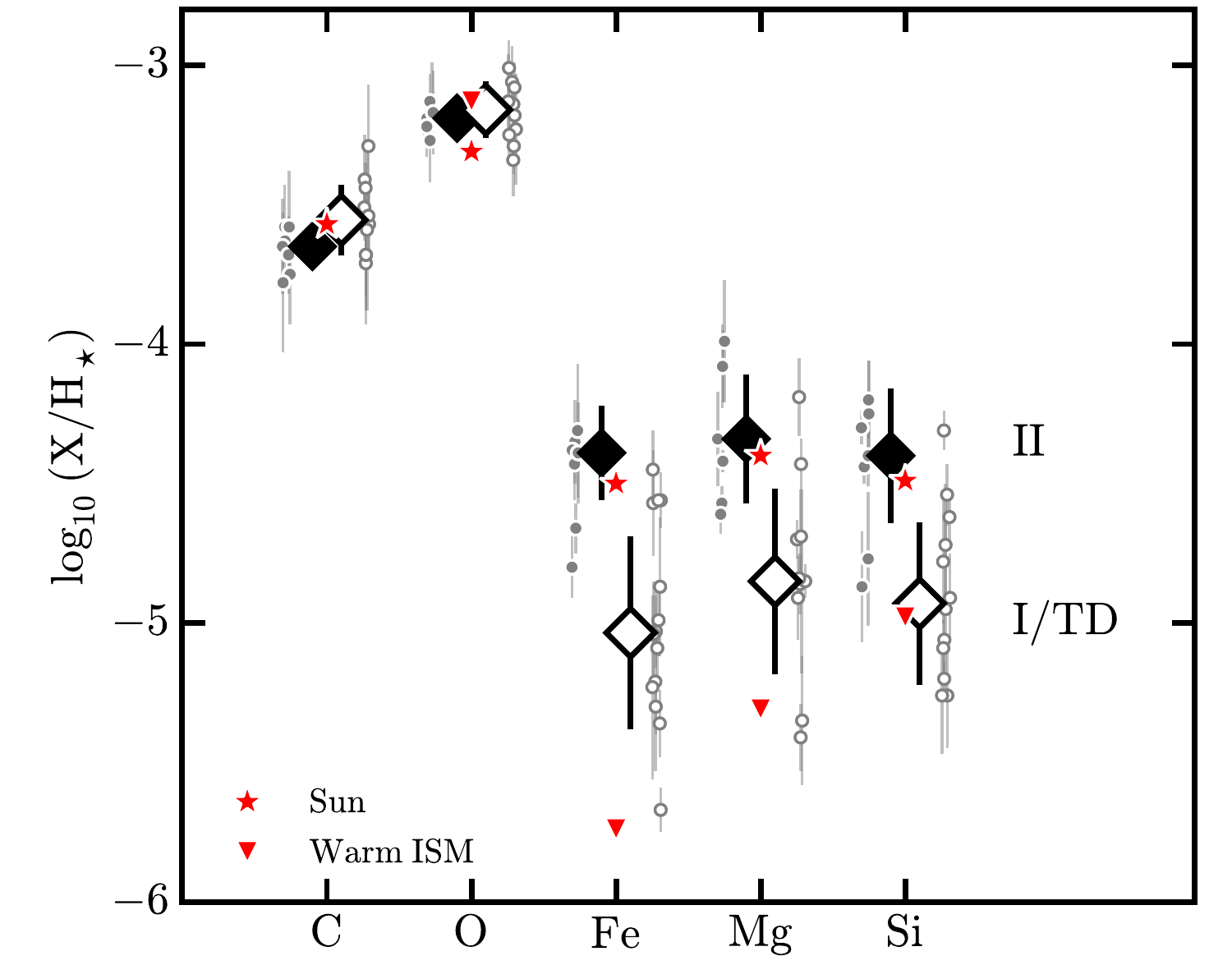}
\caption{Elemental abundances in the photospheres of Herbig stars hosting group~I (transitional/TD, open symbols) and group~II (filled symbols) disks. The sun and the warm diffuse interstellar medium are shown in red. The full Herbig samples (circles) are shown, as well as their medians with sample standard deviations (diamonds with bars).}
\label{fig:avgabuns}
\end{figure}

\revone{We adopt the photospheric abundances from \citet{AckeWaelkens2004} and \citet{Folsometal2012}. For overlapping sources, the surveys are largely consistent, but we adopt the Folsom et al. values as the study is more consistent in the stellar parameter determination. For some Acke \& Waelkens sources, we used values based on the ionization state with the largest number of analyzed lines. Some stars in the abundance studies host debris and not protoplanetary disks and were excluded. The abundances for HD~100546 are from Kama et al. (submitted) and for HD~34282 from \citet{Merinetal2004}.} HD~101412 was identified as a \lboo\ star by \citet{Cowleyetal2010}, we use newer abundances from \citet{Folsometal2012}. The solar abundances shown in Fig.~\ref{fig:avgabuns} are from \citet{Asplundetal2009}. Values for the warm diffuse interstellar gas are from \citet{SavageSembach1996}.

\subsection{Properties of the disks}

Most of the disk group classifications are from \citet{Maaskantetal2014}. The references for radial disk cavities and gaps and for stellar accretion rates, spectral types and ages are given in Table~\ref{tab:systems}. HD~101412 is a \lboo\ star with a flat, cold dust disk consistent with a group~II disk, but has a flaring, warm gas disk more reminiscent of group~I systems \citep{Fedeleetal2008, vanderPlasetal2008}. We have classified it as group~I or transitional, as it is clearly not a pure group~II source. HD~245185 is the only remaining massive dust disk in the $5\,$Myr old $\lambda\,$Ori cluster and based on CO observations has a very low gas to dust mass ratio \citep{Ansdelletal2015}. HD~179218 is a distant ($d=244\,$pc) group~I source which shows evidence for a characteristic dust emission radius larger than that of group~II sources, although a cavity has not definitively been resolved \citep{Leinertetal2004}.

\section{Results}\label{sec:results}

Within our sample, all but two group~I systems are confirmed to be transitional, with a large-scale reduction of the surface density of millimetre-size dust in the inner disk. As a working hypothesis, we consider the group~I and transitional classifications to be identical (I/TD). We do not employ a quantitative cutoff between stars without and with \lboo\ abundance patterns, although some of the stars in the sample display prominent \lboo\ anomalies. Rather, we separate the sources into group~I and II disk hosts, and compare the stellar abundances between these subsets.

Figure~\ref{fig:avgabuns} shows the photospheric abundances of two volatile elements (C and O) and three rock-forming elements (Fe, Mg and Si) for the median and full sample of group~I/TD and group~II disk hosting stars. The solar photospheric abundances and those of the warm diffuse interstellar medium are given for comparison. The median elemental abundances are summarized in Table~\ref{tab:medians}. The medians for the group~II disk host stars are solar, while the medians for the group~I/TD are solar-like for the volatiles, but depleted from solar by $0.5\,$dex for all refractory elements. \revone{The medians are unchanged if we exclude the Acke \& Waelkens sample -- all $\rm \log_{10}(X/H)$ values change by $\leq0.03$, except Mg/H which increases by $0.13$ for group~II.}

Only two group~II sources, HD~142666 and HD~144432, have strongly sub-solar refractory element abundances. Interferometric observations reveal radial gaps on (sub-)au scales in both \citep{Chenetal2012, Schegereretal2013, Menuetal2015}. They are thus small-scale transitional disks with outer disks similar to group~II sources. The group~II/I unclear source HD~101412 -- for which we adopted group~I -- has a narrow ring of CO emission near the dust sublimation zone, also indicative of small-scale radial structure \citep{Cowleyetal2012}. The disks around HD~101412, HD~142666 and HD~144432 all have large radial variations of surface density at ${\lesssim}1\,$au, with host stars depleted in refractory elements. As noted by \citet{Currie2010}, such systems may probe the earliest stages of disk clearing. 

\begin{table}[!ht]
\caption{\revone{Median $\rm\log_{10}(X/H)$ abundances and sample standard deviations (scatter within the subset) for group~I and II disk host stars.}}\label{tab:medians}
\centering
\begin{tabular}{c | c c | c c c }
\hline\hline
		& C/H$_{\star}$	& O/H$_{\star}$	& Fe/H$_{\star}$	& Mg/H$_{\star}$	& Si/H$_{\star}$	\\
\hline
II		& $-3.65$	& $-3.19$	& $-4.39$	& $-4.34$	& $-4.40$	\\
$\sigma$	& $0.07$	& $0.05$	& $0.17$	& $0.23$	& $0.24$	\\
\hline
I/TD		& $-3.57$	& $-3.14$	& $-5.04$	& $-4.84$	& $-4.95$	\\
$\sigma$	& $0.10$	& $0.10$	& $0.35$	& $0.35$	& $0.30$	\\
\hline
\end{tabular}
\end{table}

\subsection{The dust to gas ratio in the inner disk}

\begin{figure}[!ht]
\includegraphics[clip=,width=1.0\columnwidth]{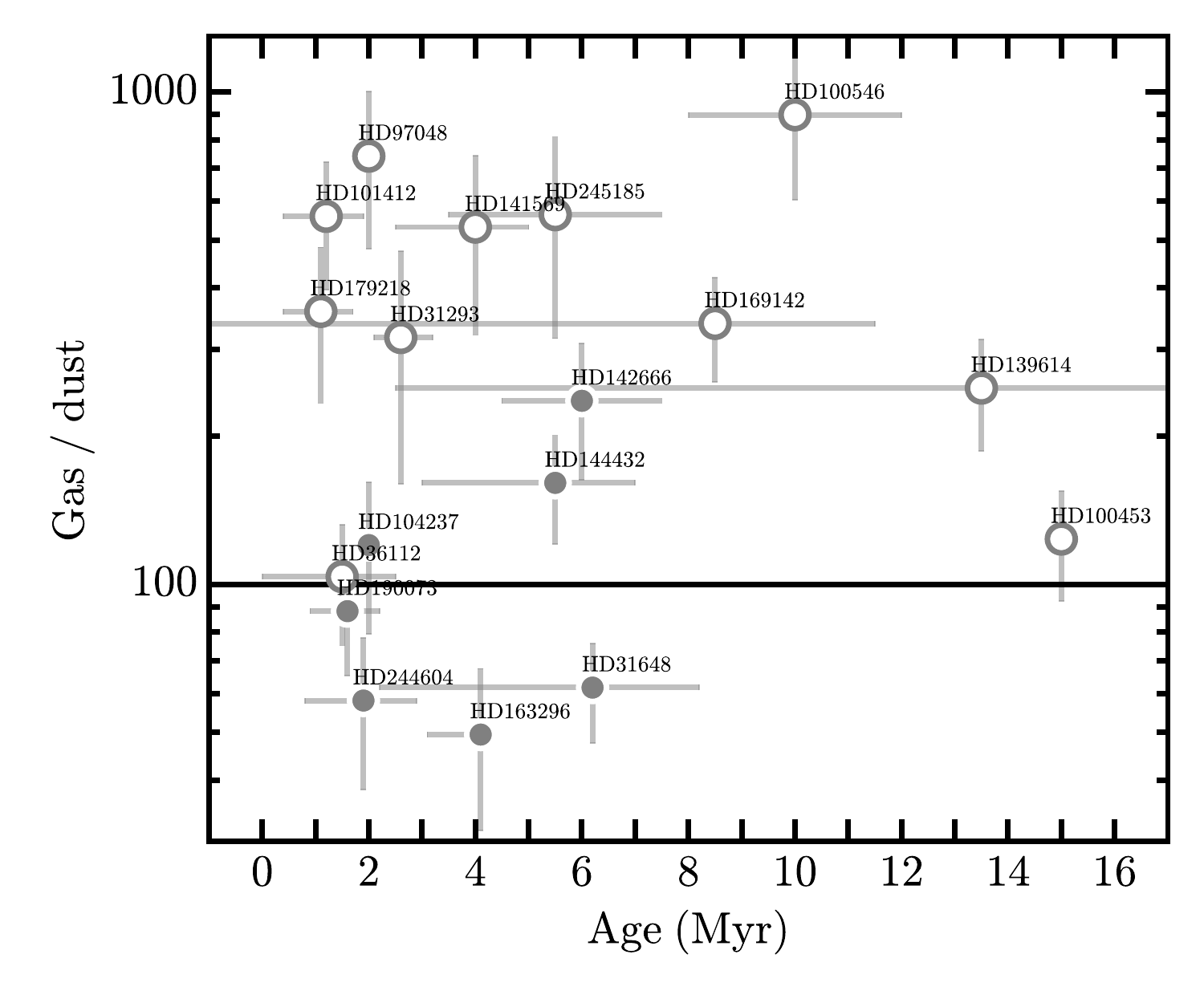}
\caption{The gas to dust ratios in the accreting material of group~I (open circles) and group~II (filled) disks, calculated from the stellar abundances with Eq.~\ref{eq:gasdust}.}
\label{fig:gasdust}
\end{figure}

Group~II disks are have no major depletion of dust or gas in their inner regions, while group~I sources have surface density reduction factors of $10^{-1}$ to $\gtrsim10^{-6}$ for dust, and none to $10^{-5}$ for gas \citep{Brudereretal2014, Zhangetal2014, Perezetal2015, vanderMareletal2015}. A gap-opening giant planet reduces the surface density of large (millimetre-scale) dust by a larger factor than that of the gas \citep{Riceetal2006, Zhuetal2012}. This offers a natural explanation for a decreased photospheric abundance of refractory elements, as a large fraction of the dust mass can remain trapped further out in the disk while gas and smaller grains continue to accrete. We now consider the photospheric composition of Herbig stars as a proxy for the gas to dust ratio in the accreting material. The solar composition corresponds to a gas to dust ratio of $100$. Noting the abundance of volatiles (C, O) as V/H and those of refractories (Fe, Mg, Si) as R/H, we express the $\Delta_{\rm g/d}$ ratio in an inner disk feeding an accretion flow as
\begin{equation}\label{eq:gasdust}
\Delta_{\rm g/d} = 100\times \left( \frac{\rm R/H_{\odot}}{\rm R/H_{\star}} \right) \left( \frac{\rm V/H_{\star}}{\rm V/H_{\odot}} \right) ,
\end{equation}
where V/H$_{\star}$ and R/H$_{\star}$ are the abundances in the host star. The results are shown in Fig.~\ref{fig:gasdust} for systems where at least one volatile and one refractory element has been measured.

\subsubsection{HD~100546}\label{sec:hd546}

The highest inner disk gas to dust ratio in our sample, $1000$, is found for HD~100546. We compare this photosphere-based value to the two-planet hydrodynamic disk model from \citet{Walshetal2014b} and \citet{Pinillaetal2015b}, and the physical-chemical disk model from Kama et al. (submitted). \revone{At $10\,$Myr, the age of the system, the hydrodynamic model yields $\Delta_{\rm g/d}\approx1000$ inside $20\,$au. While this model was not configured to model the accretion zone, the gas to dust ratio of the inner cavity provides a zeroth-order approximation for the accreting material. The model from Kama et al. reproduces the spectral energy distribution and a range of emission lines, and allows for $\Delta_{\rm g/d}=1000$ in the inner disk. The consistency of all three approaches suggests that the photospheric abundances reflect the inner disk material.}

\section{Discussion}

\subsection{Evidence for giant planets}

\begin{figure}[!ht]
\includegraphics[clip=,width=1.0\columnwidth]{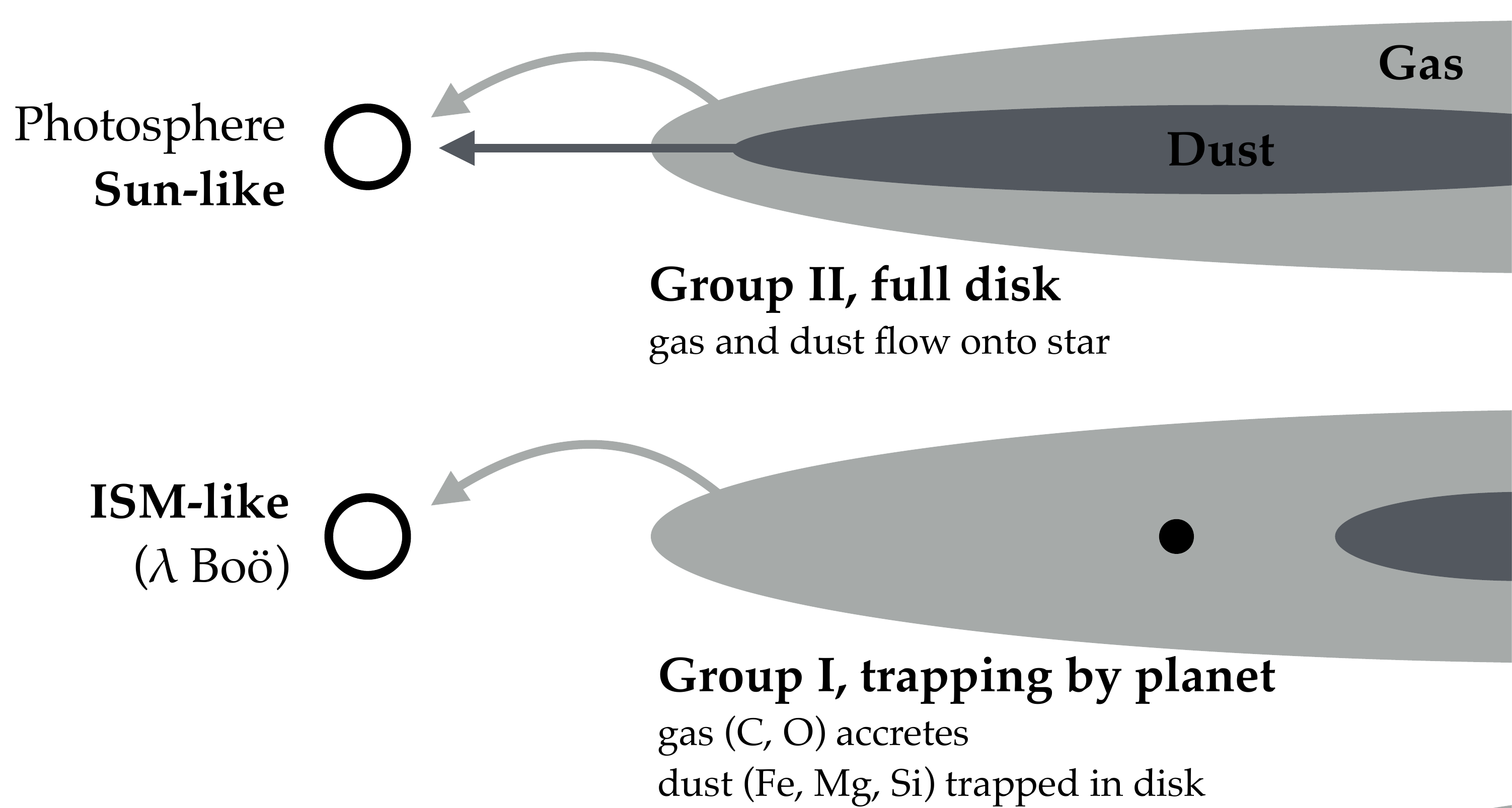}
\caption{The stellar photospheric abundances and the inner disk gas and dust distribution for group~I (transitional) and II sources.}
\label{fig:cartoon}
\end{figure}

\revone{The depletion of refractory elements in group~I host stars is consistent with giant planets interacting with their disks.} A planet with a mass $\lesssim0.1\,$\mjup\ does not substantially alter the surface density of dust or gas, with $\Delta_{\rm g/d}\approx100$ at the planet position \citep[e.g.,][]{Dongetal2015}. A planet in the mass range $1\,$\mjup$\rm \lesssim M_{p}\lesssim 10\,$\mjup\ decreases the surface density of millimetre-sized dust more than that of the gas, with $\Delta_{\rm g/d}\approx1000-10000$ \citep[][Fig.~6]{Pinillaetal2012b}. A more massive planet, $M_{\rm p}\gtrsim10-15\,$\mjup, substantially decreases the surface densities of both the dust and the gas. Since the observations require a larger decrease of the dust than the gas in the accreted material, we propose that the \lboo-like Herbig stars host planets in the ${\gtrsim}0.1$ to $10\,$\mjup\ range, with the exact limits depending on the disk temperature and viscosity. Photoevaporation would deplete gas at least as efficiently as dust and is thus not consistent with the \lboo\ sources. We illustrate the typical group~II and I inner disk structures and their relation to stellar photospheric abundances in Fig.~\ref{fig:cartoon}.

Fig.~\ref{fig:avgabuns} shows that while the underabundance of refractory elements in group~I/TD host stars is significant, not all are strong \lboo\ stars, and some have solar-like abundances. This may be related to the presence of an inner disk extending to the accretion radius \citep[``pre-transitional'' disks, see][]{Espaillatetal2014} and still containing enough mass in gas and dust to not be affected by the depletion of large grains in the gap containing the planet. Alternatively, such disks may host companions with masses $\gtrsim10\,$\mjup, which would block the accretion of both gas and dust from the outer disk. The stars in our sample have high accretion rates, which favours the massive inner disk scenario.

\subsection{The occurrence rates of transitional disks, \lboo\ anomalies, and planets}

Based on a large homogeneous study of T~Tauri stars with the SMA interferometer, \citet{Andrewsetal2011a} found an occurrence rate of transitional disks of $>26\,$\%. As a first guess, we adopt this value for the Herbig~Ae/Be systems. If the \lboo\ anomalies are due to the filtration of dust in transitional disks, we expect a similar occurrence rate of \lboo\ anomalies. Among the Herbig sample of \citet{Folsometal2012}, the \lboo\ occurrence rate is $33\,$\%, consistent with the estimated transitionality fraction.

If the correlation of refractory depletion and disk structure is indeed due to giant planets, and if the transitional phase is short compared to the disk lifetime, the $\sim30\,$\% occurrence rate of transitionality and \lboo\ anomalies is a lower limit on the fraction of Herbig stars hosting giant planets in the $0.1$--$10\,$\mjup\ range. Current exoplanet surveys are not complete down to $0.1\,$\mjup\ at Jupiter-like orbits. In the $3$--$14\,$\mjup\ mass range, the occurrence rate of planets at $5$--$320\,$au around A stars is $3$ to $30\,$\% at $95\,$\% confidence \citep{Viganetal2012}, while the ${\geq}4\,$\mjup\ planet occurrence rate at $59$--$460\,$au for B and A stars is ${\leq}20\,$\% \citep{Nielsenetal2013}. \revone{The radial velocity survey of \citet{Reffertetal2015} found an occurrence rate of ${\sim}30\,$\% for planets ${\gtrsim}1\,$\mjup\ around intermediate-mass stars. \revtwo{This peaks towards super-solar metallicity ([Fe/H]~$\gtrsim0.2$) and declines for stars more massive than ${\sim}2.5\,$\msol.} Allowing for undetected lower-mass planets, the occurrence rate of transitionality and \lboo\ anomalies is consistent with the giant planet population around A stars. \revtwo{The fact that the radial velocity planet occurrence rate peaks at a bulk stellar metallicity [Fe/H]~$\gtrsim0.2$ suggests that \lboo-like depletions typically decrease photospheric metallicities by more than $0.5\,$dex.}}

\subsection{The longevity of B, A, F star disks}

A large fraction of the disks studied in this paper are older than the canonical lifetime of $2$--$6\,$Myr for massive gas-dust disks \citep{Haischetal2001}. Several sources are in the $10\,$Myr regime. At such ages, many systems already show low emission levels consistent with debris disks \citep[e.g.,][]{Hardyetal2015, Wyattetal2015}. This is particularly surprising since A stars are thought to lose their disks faster \citep{Ribasetal2015} and suggests that the giant planet systems responsible for clearing large fractions of the inner disk may contribute to disk longevity by trapping material in the outer disk. It has been suggested that ${\sim}30\,$\% of field stars retain their gas disks for for $10\,$Myr \citep{Pfalzneretal2014}.

The frequency of the \lboo\ phenomenon among main sequence B, A and F type stars is ${\approx}2\,$\% \citep{GrayCorbally1998}. \revone{These stars likely do not host massive protoplanetary disks and their \lboo\ peculiarity should disappear on a ${\sim}1\,$Myr timescale.} However, a correlation has been found between a star showing \lboo\ peculiarities and hosting a debris disk \citep[][Draper et al., submitted]{KingPatten1992}. This suggests that debris disks around early-type stars generally retain or produce large amounts of volatile-rich gas for up to Gyr timescales. \revone{We note that HR~8799, a $30\,$Myr old intermediate-mass star which hosts four giant planets of $5$--$7\,$\mjup\ at large orbits, as well as a multi-component debris disk, is a \lboo\ peculiar star \citep{Maroisetal2008, Suetal2009, Maroisetal2010, Murphyetal2015}.} The photospheric abundances of early-type stars are a new window into the long-term evolution of protoplanetary gas disks.

\section{Conclusions}

We present a correlation between the presence of a large radial cavity or gap in a protoplanetary disk and the depletion of refractory elements in the photosphere of the host star. Such a depletion in intermediate-mass stars, known as the \lboo\ peculiarity, has been known for decades but has until now been unexplained. We propose that the trapping of large dust grains in the outer disk by giant planets provides a natural mechanism for increasing the gas to dust ratio of the accreted material, suggesting that as many as ${\sim}30\,$\% of Herbig~Ae/Be stars harbor giant planets. 

\begin{enumerate}
\item{A typical late-B through early-F star hosting a group~I or transitional disk is depleted in rock-forming elements (Fe, Mg, Si) by $0.5\,$dex compared to solar and group~II host abundances.}
\item{All group~II disk hosts in our sample have solar-like stellar abundances. The two exceptional systems, HD~142666 and HD~144432, have small-scale radial dust gaps.}
\item{The depletion of rock-forming elements in stars hosting group~I disks \revone{is likely due to} giant planets which trap large dust grains in the outer disk. The diversity of disk structures and planet masses explains why \lboo-like stars have a range of depletions, rather than clustering at particular values.}
\item{A deficiency of rock-forming elements in the photosphere of a disk-hosting B, A or F star implies the presence of a radial dust cavity or gap, and a giant planet.}
\item{Most as-yet uncharacterized central stars of transitional Herbig~Ae/Be disk systems \revone{will likely be found} to have a low abundance of rock-forming elements.}
\end{enumerate}

\begin{acknowledgements}
Astrochemistry in Leiden is supported by the Netherlands Research School for Astronomy (NOVA), by a Royal Netherlands Academy of Arts and Sciences (KNAW) professor prize, and by the European Union A-ERC grant 291141 CHEMPLAN. CPF was supported by the French grant ANR 2011 Blanc SIMI5-6 020 01``Toupies: Towards understanding the spin evolution of stars''.
\end{acknowledgements}

\bibliographystyle{aa}
\bibliography{diskstar}

\appendix

\section{Details of the sample}

\begin{sidewaystable*}[!ht]
\caption{The systems studied in this paper.}\label{tab:systems}
\begin{tabular}{ c c c c c c c c c c c c c c }
\hline\hline
System	&	Spec.	& $\log(g)$		& Age	& C/H$_{\star}$	& O/H$_{\star}$ & Fe/H$_{\star}$ & Mg/H$_{\star}$ & Si/H$_{\star}$ &	$\log_{10}$\.{\em M}$_{\rm acc}$		&	Meeus	& $r_{\rm hole}$ & $r_{\rm gap}$	& Refs.	\\
		&	type		& (cm$\,$s$^{-1}$)		& (Myr)	&	& 	& 	& 	&  	&			(M$_{\odot}\,$yr$^{-1}$)	&	group	&	(au)		& (au)	&	\\
\hline
HD~31648	& A5		& $4.1\pm0.2$		& $6.2^{+4.0}_{-2.0}$	& $3.63\pm0.07$	& $3.24\pm0.05$	& $4.43\pm0.13$ 	& $4.08\pm0.13$	& $4.25\pm0.19$	& $-6.90$		& IIa	&	&				& $^{1,7,19}$ \\	
HD~163296	& A3		& $4.2\pm0.3$		& $4.1^{+1.0}_{-0.2}$	& $3.78\pm0.25$	& $3.27\pm0.15$	& $4.35\pm0.15$	& $4.08\pm0.15$	& $4.20\pm0.14$	& $-7.16$		& IIa	&	&				& $^{1,7,19}$	\\
HD~244604	& A0		& $4.0\pm0.2$		& $1.9^{+1.1}_{-1.0}$	& $3.65\pm0.17$	& $3.19\pm0.08$	& $4.31\pm0.24$ 	& $3.99\pm0.22$	& $4.30\pm0.06$	& $-7.20$		& IIa	&	&				& $^{1,7,19}$	\\
HD~190073	& A2		& $3.7\pm0.3$		& $1.6^{+0.7}_{-0.6}$	& $3.68\pm0.14$	& $3.22\pm0.11$	& $4.38\pm0.06$ 	& $4.34\pm0.17$	& $4.44\pm0.06$	& $-8.68$		& IIa	&	&				& $^{1,7,18}$	\\
HD~104237	& A4		& $4.5$			& $2.06$				& $3.58\pm0.20$	& $3.17\pm0.15$	& $4.39\pm0.18$ 	& $4.61\pm0.07$	& $4.40\pm0.11$	& $-7.45$		& IIa	&	&				& $^{2,7,17,20}$	\\
HD~142666	& A8		& $3.9\pm0.3$		& $6.0^{+1.5}_{-1.5}$	& $3.58\pm0.15$	& $3.14\pm0.15$	& $4.80\pm0.11$ 	& $4.57\pm0.06$	& $4.87\pm0.20$	& $-7.77$		& IIa	& $0.30$	& $0.35-0.80$	& $^{1,7,9,19}$\\
HD~144432	& F0		& $3.9\pm0.3$		& $5.5^{+2.5}_{-1.5}$	& $3.75\pm0.18$	& $3.13\pm0.10$	& $4.66\pm0.09$ 	& $4.42\pm0.04$	& $4.77\pm0.24$	& $-7.74$		& IIa	& $0.18$	& $0.22-1$	& $^{1,7,10,19}$\\
HD~101412	& B9.5	& $4.0\pm0.5$		& $1.2^{+0.8}_{-0.7}$	& $3.54\pm0.12$	& $3.08\pm0.09$	& $5.04\pm0.19$ 	& $4.91\pm0.15$	& $5.26\pm0.19$	& $-7.04$		& IIa/Ia & $0.4$	&	 		& $^{1,11,12,18	}$\\
HD~31293	& A0		& $3.9\pm0.3$		& $2.6^{+0.5}_{-0.6}$	& $3.29\pm0.22$	& $3.23\pm0.20$	& $4.87\pm0.22$	& $4.85\pm0.33$	& $4.72\pm0.22$	& $-7.74$		& Ia	&	$115$	&		& $^{1,7,13,19}$\\
HD~36112	& A8		& $4.1\pm0.4$		& $1.5^{+1.5}_{-1.0}$	& $3.57\pm0.16$	& $3.14\pm0.10$	& $4.45\pm0.14$ 	& $4.19\pm0.14$	& $4.54\pm0.11$	& $-6.05$		& Ia	&	$73$	&			& $^{1,7,14,15,19}$	\\
HD~179218	& B9		& $3.9\pm0.2$		& $1.1^{+0.7}_{-0.6}$	& $3.41\pm0.16$	& $3.06\pm0.13$	& $4.99\pm0.13$ 	& $4.69\pm0.17$	& $4.78\pm0.22$	& $-6.72$		& Ia	&	Y	&			& $^{1,7,19}$\\
HD~100546	& B9		& $4.2\pm0.3$		& $10$				& $3.68\pm0.20$	& $3.25\pm0.20$	& $5.67\pm0.08$ 	& $5.41\pm0.12$	& $5.26\pm0.21$	& $-7.23$		& Ia	&	$13$	& $150-230$	& $^{3,7,16,18}$\\
HD~250550	& B9		& $4.0$		& $0.25$				& --				& --			 	& $5.36\pm0.12$ 	& $4.84$			& $4.62\pm0.09$	& $-7.80$		& Ia	&	&				& $^{2,7,17,19}$\\
HD~142527	& F6		& 		& $0.86$				& --				& --				& $4.59$ 			& --				& --				& $-7.16$		& Ia	&		& $10-140$	& $^{4,7,17,20}$\\	
HD~139614	& A7		& $3.9\pm0.3$		& $13.5^{+11}_{-5}$		& $3.71\pm0.22$	& $3.29\pm0.10$	& $5.03\pm0.13$	& $4.70\pm0.07$	& $5.06\pm0.13$	& $-7.99$		& Ia	&	$5.6$&			& $^{1,7,20}$\\
HD~34282 	& A3		& $4.2\pm0.2$		& $>7.81$				& --				& --			 	& $5.3\pm0.1$ 	& --				& --				& $<-7.71$	& Ib	&	$92$	&			& $^{5,7,17,20,21}$\\ 
HD~141569	& B9.5	& $4.2\pm0.4$		& $4.0^{+1.5}_{-1.0}$	& $3.59\pm0.29$	& $3.01\pm0.10$	& $5.21\pm0.32$	& $4.86\pm0.11$	& $4.95\pm0.31$	& $-8.37$		& Ib	&	$95$	&			& $^{1,5,7,20}$\\	
HD~169142	& A5		& $4.3\pm0.2$		& $8.5^{+16}_{-3}$		& $3.51\pm0.12$	& $3.34\pm0.13$	& $5.09\pm0.11$	& $4.85\pm0.06$	& $5.09\pm0.10$	& $-7.40$		& Ib	&	$23$	&$40-70$		& $^{1,6,7,20}$\\	
HD~100453	& A9		& $4.0$		& $10$				& $3.44\pm0.09$	& --				& $4.57\pm0.19$ 	& $4.43\pm0.09$	& $4.31\pm0.07$	& $-8.04$		& Ib	&	$20$	&			& $^{2,7,17,20,21}$\\		
HD~97048 	& A0		& $4.0$		& $1.37$				& --				& $3.18\pm0.14$	& --			 	& --				& $5.26\pm0.06$	& $-6.80$		& Ib	&	$34$	&			& $^{2,6,7,17,18}$\\	
HD~135344~B & F8		& $4.0$		& $8.93$				& --				& --				& $4.56\pm0.10$	& --				& --				& $-7.69$		& Ib	&	$40$	&			& $^{2,7,17,18}$	\\	
HD~245185	& A0		& $4.0\pm0.4$		& $5.5^{+2.0}_{-2.0}$	& $3.68\pm0.15$	& $3.13\pm0.17$	& $5.23\pm0.33$	& $5.35\pm0.23$	& $4.91\pm0.16$	& $-7.20$		& I	&			&		& $^{1,8,19}$\\	
\hline
Sun 			& G2		& $4.4$		& $4600$				& $3.57\pm0.05$	& $3.31\pm0.05$	& $4.50\pm0.04$ 	& $4.40\pm0.04$	& $4.49\pm0.03$	&  --		&  --	& --	& --				& $^{22}$	\\
\hline
\end{tabular}
\flushleft
\emph{Notes. }Stellar elemental abundances are given here as $\rm -\log_{10}{\left( X/H_{\star} \right)}$ (note the minus), \revone{while \citet{Folsometal2012} give $\rm \log_{10}{\left( X/{total}_{\star} \right)}$ in their Table~4}. Hole and gap radii refer to those seen in millimetre-sized grains. All uncertainties are $1\,\sigma$. `Y' denotes cases where inner cavities have been inferred, but not well quantified, based on evidence complementary to the spectral energy distribution.\\
\emph{References. }
1 -- \citet{Folsometal2012};
2 -- \citet{AckeWaelkens2004};
3 -- Kama et al. (submitted);
4 -- \citet{Holmbergetal2009};
5 -- \citet{Merinetal2004};
6 -- \citet{Maaskantetal2013};
7 -- \citet{Maaskantetal2014};
8 -- \citet{Ansdelletal2015};
9 -- \citet{Schegereretal2013};
10 -- \citet{Chenetal2012};
11 -- \citet{Fedeleetal2008};
12 -- \citet{vanderPlasetal2008};
13 -- \citet{Pietuetal2005};
14 -- \citet{Isellaetal2010};
15 -- \citet{Andrewsetal2011a};
16 -- \citet{vandenAnckeretal1997};
17 -- \citet{Manojetal2006};
18 -- \citet{Pogodinetal2012};
19 -- \citet{DonehewBrittain2011};
20 -- \citet{GarciaLopezetal2006};
21 -- Khalafinejad et al. (submitted);
22 -- \citet{Asplundetal2009}
\end{sidewaystable*}

\end{document}